\documentstyle[prc,aps,preprint]{revtex}
\begin{document}
\draft
\title{Electron screening
in molecular fusion reactions}
\author{T.~D.~Shoppa$^1$, M. Jeng$^1$, S.~E.~Koonin$^1$, 
K.~Langanke$^1$, and R.~Seki$^{1,2}$}
\address{$^1$ W.~K.~Kellogg Radiation Laboratory, 106-38\\
California Institute of Technology\\
Pasadena, CA 91125 USA\\
$^2$Department of Physics and Astronomy\\
California State University\\
Northridge, CA 91330}
\date{\today}
\maketitle

\begin{abstract}
Recent laboratory experiments have measured fusion cross
sections at center-of-mass energies low enough for the
effects of atomic and molecular electrons
to be important. To extract the cross section for
bare nuclei from these data (as required for astrophysical applications),
it is necessary to understand these screening  
effects. 
We study electron screening effects in the low-energy collisions
of $Z=1$ nuclei with hydrogen molecules. Our model is based 
on a dynamical evolution of the electron  
wavefunctions within
the TDHF scheme, while the motion of the  
nuclei
is treated classically. 
We find that at the currently accessible energies the screening effects
depend strongly on the molecular orientation. 
The screening is found to be
larger
for molecular targets than for atomic targets,
due to the reflection symmetry in the latter.
The results agree fairly well with data measured for deuteron
collisions on molecular deuterium and tritium targets.
\end{abstract}

\section{ Introduction and motivation}

The determination of nuclear cross sections at the astrophysically most 
effective energies requires in most cases an extrapolation of cross
sections that have been measured 
at the lowest possible laboratory energies.
It has been generally believed that the uncertainty in this
extrapolation is reduced by lowering the energies at which data
can be taken in the laboratory. However, Assenbaum {\it et al.}
\cite{Assenbaum} pointed out that this strategy might be problematic
as, at very low energies, the experimental cross sections
do not reflect the collisions of bare nuclei. 
Rather, they are larger due
to screening 
by the electrons present in the target
and projectile. This screening has to be accounted for in
a determination of the nuclear cross section required for 
astrophysical applications.

Screening by the target electrons 
has been observed in several low-energy
fusion experiments
(\cite{NMR}
and references therein). Conventionally,
the effects of electron screening
are expressed in terms of an enhancement factor
\begin{equation}
f(E) = \sigma_{\rm exp} (E) / \sigma_{\rm bare} (E)
\end{equation}
where $\sigma_{\rm bare}$ is the cross section for bare nuclei
at center-of-mass energy $E$,
and $\sigma_{\rm exp}$ is the cross section measured in the presence
of target and projectile electrons. 
The experimental analysis 
generally assumes that the electrons
can be thought of as 
effectively lowering the Coulomb energy between the two colliding nuclei
by a {\it constant and energy-independent} energy increment $U_e$,
the {\it screening energy}. 
Furthermore it is assumed that the bare-nuclei cross section (usually
derived by extrapolation of data at higher energies, which are little
affected by screening) is known.
Realizing that $U_e << E$ at those energies
currently accessibly in the laboratory,
the enhancement factor can be written as \cite{Assenbaum}
\begin{eqnarray}
f(E) &=& {\sigma_{\rm bare}(E+U_e) \over \sigma_{\rm bare}(E)} =
{S(E+U_e) \over S(E)} {E \over E+U_e}
{\exp(-2\pi\eta(E+U_e)) \over \exp(-2\pi\eta(E))} \nonumber \\
& \approx & \exp \left\{ \pi \eta(E) \frac{U_e}{E} \right\}  .
\end{eqnarray}
Here, as is conventional in nuclear astrophysics,
 we have written the cross section
in terms of the astrophysical $S$-factor
\begin{equation}
\sigma(E) = {S(E) \over E} \exp (-2\pi\eta(E))\;,
\end{equation}
where
$\eta(E) = Z_1 Z_2 \alpha (\mu c^2 / 2 E)^{1 \over 2}$
is the Sommerfeld parameter for
nuclei of charges $Z_1, Z_2$ and reduced
mass $\mu$. As is obvious from the exponential energy dependence,
electron screening effects rapidly become more important with decreasing
collision energy.

The picture underlying the experimental analysis of electron
screening effects has been confirmed in recent theoretical studies for 
{\it atomic targets} \cite{Bracci,Shoppa}.
In fact, these calculations show that, at those energies at which screening
effects have been observed for atomic targets, the effective screening
potential is nearly constant at distances smaller than the classical
turning point and can be replaced by a 
spatially constant screening energy.
Furthermore it has been found that, to a good approximation, the electrons
can be treated adiabatically at these low energies, thus allowing one
to replace the screening energy by the gain in binding energy
between the united atom and the asymptotically separated fragment atoms.
However, it is as yet unexplained why the experimentally determined
screening energies apparently exceed the adiabatic limits 
in some cases\cite{NMR}.

A recent series of impressive experiments
\cite{NMR,Engstler,Angulo,Prati} has established
that the screening energy depends on the
form of the target (atomic, molecular, solid, etc.). 
In particular, it has been reported that, for the same
nuclear reaction, the electron screening is generally smaller
for molecular targets than for atomic targets. This has been attributed
\cite{Prati} to the fact that, in the case of molecular targets, the gain
in electron binding is reduced by the energy spent to break the
molecular bond and by the energy transferred to the spectator nucleus
in the molecule. Although 
this dynamical picture qualitatively explains the observed difference
between molecular and atomic targets, it has not 
yet been tested theoretically. 

To study screening in molecular targets,
we have performed calculations of collisions of $Z=1$ nuclei
with hydrogenic molecules. In our model the wave functions of the two electrons
are evolved dynamically within the time dependent Hartree-Fock (TDHF) scheme,
while the motion of the nuclei is treated classically.
That is, we  
assume
that 
the colliding nuclei behave as
classical particles 
beyond the classical turning point, 
subject to the forces they exert on  
each other
and to the force exerted on them by the electrons and the spectator nucleus.
However, the electrons are treated quantum mechanically in
the TDHF scheme, where
the wave functions are calculated
in the time-dependent potential generated by the nuclei
and the inner-electron repulsion. In this way, we  
determine the value of
the screening potential $U(R)$ between the two colliding nuclei.
Additionally we monitor the force exerted on the spectator nucleus by
the other charges to test the dynamical picture underlying
the previous analysis of electron screening effects for molecular targets.

Our paper is organized as follows. In Section 2, we
briefly describe the TDHF method applied in our calculation, as well  
as
the numerical methods involved. Our results are presented and  
discussed
in Section 3.

\section{The TDHF method}

We consider the collision of a $Z=1$ nucleus
with a $Z=1$ target nucleus bound in a hydrogenic molecule. 
The nuclei can be treated as point-like
and the collision as head-on, since the
nuclear interaction occurs at distances far smaller than 
atomic scales. Our choice
of nuclear coordinates is defined in Fig. 1. Initially, the electron 
distribution
is cylindrically symmetric about the axis defined by the two 
molecular nuclei
($z$-axis). This symmetry is broken as the  
projectile nucleus approaches.
We characterize the trajectory of the projectile nucleus by the 
orientation angle
$\theta$ between the initial projectile velocity 
and the molecular axis. In our notation
a collision where the projectile passes very near to the spectator
nucleus is described by $\theta \approx 180^\circ$.

With our choice of coordinates the Hamiltonian for the problem reads
\begin{equation}
H=
\sum_{i=1,2} \frac{p_i^2}{2 m} +
\sum_{j=p,s,t} \frac{P_j^2}{2 M_j}
+V_{pt}+V_{ps}+V_{st}+V_{en}+V_{ee}
\end{equation}
where $m$ is the electron mass and the $M_j$ are the masses of the
projectile (p),
target (t) and spectator (s) nuclei.  $V_{pt}$, $V_{st}$, and
$V_{ps}$ denote the various Coulomb repulsions among the nuclei. 
The Coulomb interaction
between electrons (labeled $i=1,2$) and the nuclei is
\begin{equation}
V_{en} = - \sum_{i=1,2} \sum_{j=p,s,t} \frac{e^2}{|{\bf r}_i - {\bf R}_j|} 
\end{equation}
and
$V_{ee}=e^2/|{\bf r}_1-{\bf r}_2|$ is the interaction between
the two electrons.

The Hamiltonian for the isolated hydrogen molecule is given by
\begin{equation}
H=
\sum_{i=1,2} \frac{p_i^2}{2 m} +
\sum_{j=s,t} \frac{P_j^2}{2 M_j} +
V_{st}+V_{ee}-
\sum_{i=1,2} \sum_{j=s,t} \frac{e^2}{|{\bf r}_i - {\bf R}_j|} .
\end{equation}

In the Hartree-Fock approximation to the
ground state,
the electronic wave function can be written  
as
\begin{equation}
\Psi({\bf r}_1,{\bf r}_2,t) =
 \psi({\bf r}_1,t) \psi({\bf r}_2,t)  
{1\over\sqrt{2}}(\alpha(1)\beta(2)-\beta(1)\alpha(2))\;.
\end{equation}
This wave function describes two electrons in a spin-singlet state
($\alpha$ and $\beta$ are the one-electron spin states). As our Hamiltonian
is independent of the spin coordinates, the spin structure of the
electron wave function remains unchanged during the collision.

The initial ground state of the molecule was constructed by evolving a
trial wavefunction with a fixed nuclear separation $R$
in imaginary time for the molecular Hamiltonian $H_{mol}$  
\cite{Kulander,Koonin1}, which 
damps out the highest energy components of the wave function.
The resulting HF ground state is the lowest eigenfunction of
the (discretized) Hamiltonian.
As long as the overlap between the ground state and the trial  
wavefunction
is nonzero, the wave function converges to the
Hartree-Fock ground state. 
The nuclear separation has been treated as a variational
parameter. We find the energy minimum at $R=0.78 {\rm \AA}$,
where the converged energy
is -30.8 eV. Both values are in good agreement
with experiment.

We evolve the spatial wave function $\psi$ by solving the TDHF
equation
\begin{equation}
i {\partial \psi({\bf r},t) \over \partial t }=-
\bigg(\Phi({\bf r},t) + \nabla^2 \bigg) \psi({\bf r},t)
\end{equation}
with
\begin{equation}
\Phi({\bf r},t) =  {e^2 \over |{\bf r} - {\bf R}_s(t)|} +
 {e^2 \over |{\bf r} - {\bf R}_t(t)|} +
 {e^2 \over |{\bf r} - {\bf R}_p(t)|} -
 e^2 \int d^3 {\bf r}^\prime
 {|\psi({\bf r}^\prime,t)|^2 \over |\bf{r} - {\bf r}^\prime|}   .
\end{equation}
The electron-electron potential has been calculated by solving the
appropriate Poisson equation  subject to a
spherical harmonics expansion around the center of electron charge as the boundary 
condition. The TDHF equations have been solved 
on a three-dimensional grid
using a second-order expansion in $\Delta t$. 
The uniform Cartesian grid
(with separation 0.26 ${\rm \AA}$)
had (20,20,40) points in ($x,y,z$) directions, and $\psi$ was assumed to 
vanish at the boundaries.

In accordance with our classical treatment of the nuclei, the time  
dependence of the nuclear position vectors ${\bf R}_i(t)$ is  
determined from Newton's law, where the force on each nucleus is the  
sum of the Coulomb force of the other nucleus and the force due to  
the electronic charge density.

\section{Results and discussion}

The electron response of the target
depends on the velocity of the incoming projectile.
We have studied collisions of a $Z=1$ nucleus with a hydrogenic
molecule for various collision angles $\theta$ and for projectile velocities 
between $0.2 \alpha c$ and $10 \alpha c$,
corresponding 
to center-of-mass bombarding
energies $E$ in the $d+{\rm D}_2$ system
between $E=1$ keV and 2.5 MeV.
In our discussion below we will refer to the velocity dependence of
the collision in terms of the equivalent
center-of-mass 
collision energy $E$ for the $d+{\rm D}_2$ system; for application
to other hydrogenic reactions of interest, such as $d+{\rm T}_2$,
these energies must be scaled according to the reduced mass
of the colliding nuclei.
In the calculation the projectile was aimed directly at the target nucleus
in the molecule, with an initial separation
of 6 ${\rm \AA}$.
We then followed the collision until the projectile had approached the target
to the classical turning point.

In Figs. 2-4 we show ``movies'' of collisions at $E=2.5$ MeV, $25$ keV, and 
1 keV at two selected angles, $\theta =57^\circ$ and $170^\circ$.
The time evolution of the electron wave functions is represented by contour
plots of electron densities, $2 |\psi ({\bf r},t)|^2$, projected
onto the scattering plane spanned by the three nuclei.
The positions of the nuclei at the various times were indicated by dots. In all
cases, the initial electron configuration 
is the ground state of the hydrogen molecule.

At $E=2.5$ MeV, the colliding nuclei
move much faster than the average electron velocity. In this {\it sudden limit},
the projectile falls through the potentials
generated by the electrons and the spectator
nucleus, adding the respective potential energies to the collision energy.
As expected, the electron densities 
at the classical turning point 
are independent of angle
(Fig. 5).
In the sudden limit, the collision energies are so high that
electron screening has no influence on the nuclear process.

In the adiabatic limit, the target and projectile approach each other
with a velocity much less than the average electron velocity. The electrons
are thus able to respond nearly instantaneously to the nuclear motion
and occupy the energetically most favorable configuration 
during the collision.
This situation is illustrated by the collision  
at $E=1$ keV. Upon reaching the classical turning point, the electron
configuration corresponds to the ground state of a system of charges,
$Z=2$ and $Z=1$, at 
the appropriate molecular separation. Fig. 6 shows the electron
configurations at the classical turning point for various collision angles.
Close inspection shows that the electron configurations 
vary with
angles, so that the electrons are not entirely adiabatic
even at this low collision energy (see below).

At $E=25$ keV (corresponding to a projectile velocity of $\alpha c$) 
the nuclear and electronic motions
are comparable and the electron configuration is strongly
distorted as the projectile passes through. Furthermore, the
electron response depends strongly on the collision angle,
as can be seen in Fig. 7.

During the collision we have monitored the 
expectation value of the forces exerted on the three
nuclei by the other particles
\begin{equation}
{\bf F}_p(t) 
= \int d^3 {\bf r}^\prime
{ \rho({\bf r}^\prime,t) 
[({\bf R}_p(t) - {\bf r}^\prime) ]
 \over |{\bf R}_p(t) - {\bf r}^\prime|^3 }
- \frac{e^2 ({\bf R}_p(t) -{ \bf R}_t(t)) }
      {|{\bf R}_p(t) -{ \bf R}_t(t) |^3}
- \frac{e^2 ({\bf R}_p(t) -{ \bf R}_s(t)) }
      {|{\bf R}_p(t) -{ \bf R}_t(t) |^3}
\;,
\end{equation}
with similar definitions for ${\bf F}_t, {\bf F}_s$. 
The relative force between projectile
and target, 
that is exerted by the electrons and
the spectator nucleus,
along the collision trajectory then is 
\begin{equation}
{\bf F}_{\rm rel} (t)  = 
({\bf F}_p (t) - {\bf F}_t (t)) \cdot 
\frac{ {\bf R}_p(t) - {\bf R}_t (t)}
     { |{\bf R}_p(t) - {\bf R}_t (t)|}  =
({\bf F}_p (t) - {\bf F}_t (t)) \cdot \frac { {\bf R}_{\rm rel} }
{|{\bf R}_{\rm rel}|}  \; ,
\end{equation}
Upon integration of $F_{\rm rel}$ along the collision trajectory we have
determined the molecular screening potential $U_{\rm mol}$, 
induced by the electrons
and the spectator nucleus,
\begin{equation}
U_{\rm mol} =   \int {\bf F}_{\rm rel} \cdot d{\bf R}_{\rm rel}.
\end{equation}
   
As an example, Fig. 8 shows the relative force and the screening potential
for the collision at $E=25$ keV and at various angles. Depending on the
relative dominance of the electrons and the spectator nucleus, the relative
force is either attractive or repulsive, leading to an oscillatory
behavior as a function of target-projectile separation.
At large angles, the projectile passes through the repulsive Coulomb field
of the spectator nucleus on its collision trajectory.
In these cases, the exerted relative force is repulsive at a few atomic units.
As expected, the effect is strongest at very large angles. After passing the
spectator nucleus, the electron cloud dominates; the relative force
becomes attractive again. When projectile and target are very close,
the relative force must vanish. 

The molecular screening potential is attractive in most cases, enhancing
the fusion of projectile and target. However, at large angles,
the
screening potential can also become repulsive when the projectile passes
the spectator nucleus.

As in \cite{Shoppa} we interpret the screening potential
at the classical turning point ($R_{cl}) $ as the 
(negative) screening energy, i.e.
($U_e (E,\theta)= - U_{\rm mol} (R_{cl},E,\theta)$), 
which represents the net effect of
spectator nucleus and electrons on the nuclear fusion process; i.e., we assume
that the nuclei fuse with an effective energy $E+U_e$. Fig. 9 shows
the screening energy as a function of collision energy 
at various angles.

The screening energies show a remarkable dependence on the scattering angle.
At forward angles, $U_e (E,\theta)$ increases monotonically with
decreasing collision energy. However, at back angles the screening energy
shows a minimum near $E=25-30$ keV. At extreme back angles
$(\theta \geq 160^\circ$) the minimum value even becomes negative, resulting
in ``anti-screening'' (or depletion) of the fusion cross section
under these kinematical conditions. 

At low energies,
the screening potential is attractive at all angles, and for
$E \leq 3$ keV the electronic motions becomes adiabatic and
$U_e(\theta)$ 
becomes nearly energy independent.
However, even at these low energies 
its value
depends noticeably on the molecular orientation. The
screening energy is largest for
$\theta \approx 90^\circ$, while it becomes smallest
at large angles.
This angle-dependence of the screening energy 
can be understood by considering the effect 
on the colliding nuclei
of the electron density between
the spectator and the target. 
At $\theta=90^\circ$ this electron density induces an attractive force
between projectile and target. This force becomes less attractive with
increasing or decreasing $\theta$ as the force component induced in the
$z$-direction acts like a tidal force, actually hindering the fusing 
particles.

In our discussion we have
ignored the rotation of the target molecule during the collision.
This assumption is valid,
as the angle $\Delta \theta$ that the molecule 
(with rotational energy $E_{\rm rot}$)
would rotate in the
time it takes for the projectile (with kinetic energy $E$)
to pass through the molecule is small.  In particular,
\begin{equation}
\Delta \theta \sim \sqrt{ E_{\rm rot} \over E }
\end{equation}
is only 0.005 radians in the most extreme case discussed here, where
$E$ is 1 keV and 
$E_{\rm rot}$ is 0.025 eV.

For each screening energy $U_e (E,\theta)$
one can determine a corresponding enhancement factor
$f (E, \theta)$ using Eq. (2).
For a comparison with experiment, the enhancement factor
has to be averaged over angle:
\begin{equation}
{\bar f} (E) = \frac{1}{2} \int_{-1}^{1} 
f(E,\theta) d \cos (\theta).
\end{equation}
We have performed this average by 5-point Gaussian quadrature.
Again using relation (2), ${\bar f(E)}$ can be expressed in terms
of an  average screening energy, ${\bar U_e (E)}$, which
is shown in Fig. 10.
As the extreme backward angles have only little weight in the average,
${\bar U_e}$ does not have a minimum
at $E=25-30$ keV as one might expect from $U_e(\theta)$, 
but is a monotonically decreasing function
of energy which changes from the sudden to the (nearly) adiabatic regime
between 30 keV and 3 keV.

Fig. 10 also shows the screening energy calculated for a deuteron
colliding with an atomic D-target. The results are taken from the TDHF
calculation of Ref. \cite{Shoppa}. We observe that the screening
energy at low collision energies is significantly larger for the
${\rm D}_2$-target than for the atomic ${\rm D}$-target. 
We will argue that this
is exceptional. 
It is the result of the reflection symmetry of the d+D system.
Asymptotically (large separations) the d+D wave function is a $50\%$
mixture of the gerade (positive parity) and ungerade (negative parity)
configurations \cite{Bracci}. As parity is conserved during the collision,
the united atom configuration in the adiabatic limit is then
$1/2$ (He$^+$(1s)+He$^+$(2p)), corresponding to a gain in electronic
binding energy of 1.5 Ry during the collision. As this symmetry does not
hold for the collision with the molecular target, the gain is larger and
in the adiabatic limit is approximately given by 3 Ry, corresponding
to an electron configuration similar to the one of He$^+$(1s).

More interestingly, Fig. 10 shows that the screening energy is larger
in the atomic case than in the molecular case for energies $E\geq 20$ keV.
Obviously this is due to the fact that upon angle-averaging the screening
in the molecular case is depleted due to the ``anti-screening'' effect
when the projectile has to pass the spectator nucleus
along its collision trajectory. We believe that this result is
the main difference between collisions with molecular and atomic targets
and will hold more generally. One might assume that the collision with the
molecular target at small $\theta$ roughly imitates the collision on an
atomic target. We then expect that,  in general for colliding systems
of non-identical charges the screening energy for an atomic target
is larger than for a molecular target (see Fig. 9), in agreement with the
observation reported in \cite{Prati}. We note again that the reflection symmetry
in the atomic d+D system invalidates our general argument.  

In our calculations we have also examined the force on the
spectator nucleus. 
As an example, we consider
the collision at $E=25$ keV and at $\theta=170^\circ$. We then
calculate a gain of kinetic energy of the spectator nucleus of 
$\Delta E_{\rm kin}= |\Delta {\bf P}|^2/(4 M)\approx 0.1$ eV
(in the lab frame), where $\Delta {\bf P} = \int {\bf F}_s(t) dt$ 
is the momentum transfer to the spectator.
This gain in kinetic energy is so small
that the
spectator does not move {\it before the nuclear fusion process happens}.
We have verified that this is
the case 
for all  conditions studied here. 
Thus, we do not support
the assumption that the breaking of the molecular bond and the
kinetic energy transfered to the spectator decreases the screening energy
in a collision with a molecular target
relative to an atomic target \cite{Engstler,Prati,Greife}.
We also find that the assumption that the projectile moves on
a straight line trajectory is well justified for all orientations
appearing in the Gaussian quadrature of Eq. (12).

In Ref. \cite{Greife} the $d(d,p)$ and $d(d,n)$ reactions have been studied
down to $E=1.6$ keV using a molecular target. Assuming a constant screening
energy, the value $U_e =25\pm5$ eV has been deduced from the data 
\cite{Greife} between 1.6 keV and $\approx 15$ keV. As can be seen 
in Fig. 10, our calculation
does not support the assumption of a constant screening energy in this 
energy regime, but predicts that ${\bar U_e}$ changes from $\approx 20$ eV
at $E=15$ keV to ${\bar U_e}=40$ eV at $E=2$ keV. Interpreting the screening
energy deduced in Ref. \cite{Greife} as an averaged value for the 
energy interval studied, 
our calculation appears to be in reasonable agreement with data.
Fig. 11 compares our prediction for the enhancement of the $d+D$ cross section
with the data (using Eq. (2) and our calculated ${\bar U}_e(E)$), 
assuming the same parameterized form of the bare-nuclei
S-factor \cite{Hale} as in Ref. \cite{Greife}.

Brown {\it et al.} \cite{Brown} have measured the low-energy $t(d,n)\alpha$
cross section with high precision using a $T_2$ target. In Ref. \cite{LR}
it has been argued that the cross section at energies $E \leq 16$ keV
is enhanced due to screening effects. With this assumption and
parameterizing the data for $E=16-70$ keV (which are not influenced
by electron screening), by a Breit-Wigner resonance formula, the authors
of \cite{LR} were able to consistently describe the Los Alamos data \cite{Brown}
at all energies $E\leq 70$ keV, a task which previously failed without
incorporation of screening effects. In Fig. 12 we in fact demonstrate
that the $t(d,n)\alpha$ data are well described by adding the presently
calculated screening energies to the parameterized bare nuclear S-factor,
as given in \cite{LR}.

In conclusion, we have studied the collisions of hydrogen nuclei on
hydrogenic molecules in a model that evolves the electron distributions
dynamically within the TDHF method. Our particular interest has been 
the screening effects of the electrons on the two fusing nuclei. We have
investigated this effect by determining the molecular screening
potential, induced by the electrons and the spectator nucleus, on the two
colliding nuclei. As in a previous study of collisions on atomic targets
\cite{Shoppa}, we defined the screening energy as the negative
value of the molecular
screening potential at the classical turning point and assumed that
the screening energy is added to the collision energy of the fusing particle.
We found that the screening energy depends strongly on the molecular
orientation. It is generally smaller if the projectile had to pass
near the spectator nucleus on its collision trajectory.
For the present case of collisions of $Z=1$ nuclei with a hydrogenic molecule
we found that the screening energy is larger
for the molecular target than for the atomic target. 
This is caused
by the reflection symmetry in the latter system and we do not expect
it to hold in the general case. 
We are planning to investigate this conjecture
in future studies of helium collisions on hydrogenic targets.

The work was supported in part by the National Science Foundation,
Grant Nos. PHY94-12818 and PHY94-20470, and the U.S. DOE
at CSUN, Grant No. DE-FG03-87ER40347.

\begin{figure}
\caption{
Coordinates used in our study. The $z$-axis is defined by the intra molecular
separation and $\theta$ is the angle between ${\hat z}$ and
the target-projectile separation.
}
\end{figure}

\begin{figure}
\caption{Time evolution of the electron density 
for the collision with a hydrogen nucleus on a hydrogenic molecule
at a collision energy of $E=2.5$ MeV and at $\theta=57^\circ$ (left side)
and $\theta=170^\circ $ (right side). The electron density is plotted
as equally spaced contours. The positions of the nuclei are marked
by dots. }
\end{figure}

\begin{figure}
\caption{\hbox to\hsize{Similar to Fig. 2, but for $E=25$ keV.\hfill}}
\end{figure}

\begin{figure}
\caption{\hbox to\hsize{Similar to Fig. 2, but for $E=1$ keV.\hfill}}
\end{figure}

\begin{figure}
\caption{Angular dependence of the electron density when target and projectile
are separated by a distance corresponding to the classical turning point.
The collision energy is $E=2.5$ MeV.}
\end{figure}

\begin{figure}
\caption{\hbox to\hsize{Similar to Fig. 5, but for $E=25$ keV.\hfill}}
\end{figure}

\begin{figure}
\caption{\hbox to\hsize{Similar to Fig. 5, but for $E=1$ keV.\hfill}}
\end{figure}

\begin{figure}
\caption{Relative force (in eV/${\rm \AA}$, dashed line) 
and molecular screening potential
(in eV, solid line) for a collision of $E=25$ keV and at selected angles.}
\end{figure}

\begin{figure}
\caption{\hbox to \hsize
{Screening energy as a function of collision energy and at
various angles.\hfill}}
\end{figure}

\begin{figure}
\caption{Comparison of the angle-averaged 
molecular screening energy ${\bar U}_e (E)$ (solid line)
with the atomic screening energy (dashed line, from Ref. \protect\cite{Shoppa})
as a function of collision energy.}
\end{figure}

\begin{figure}
\caption{Comparison of the experimental $d(d,p){}^3$H 
S-factors \protect\cite{Greife}
with the bare-nuclei S-factor from Ref. \protect\cite{Hale}
(dashed line) and our estimate in which electron
screening is added to the bare-nuclei S-factor using Eq. (2).}
\end{figure}

\begin{figure}
\caption{Same as Fig. 11, but for the $t(d,n)\alpha$ reaction.
The experimental data are from \protect\cite{Brown}, while the
bare-nuclei S-factor is adopted from \protect\cite{LR}.}
\end{figure}

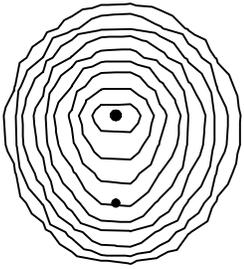
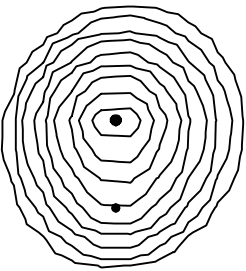
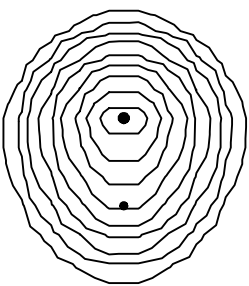
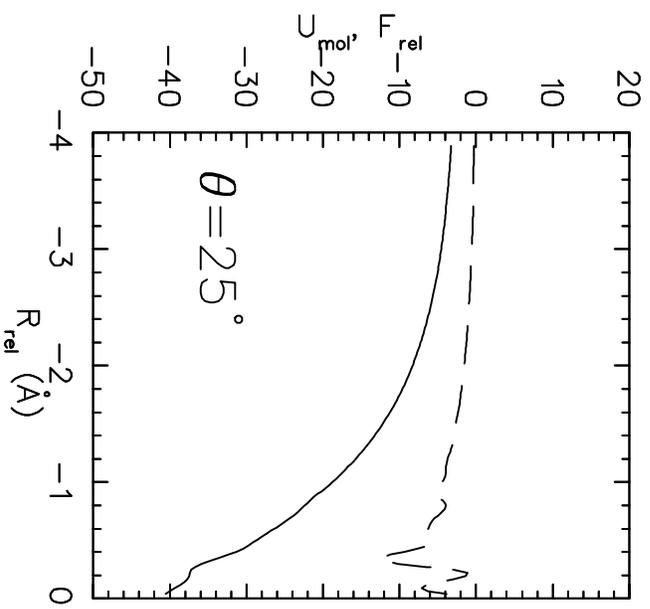
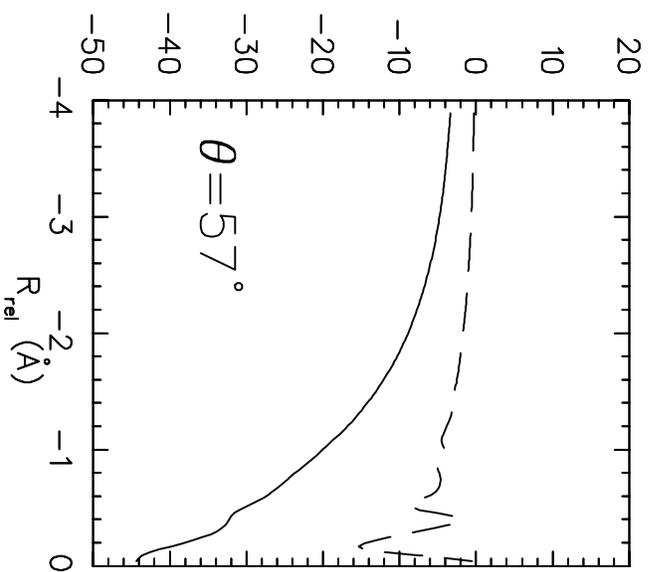
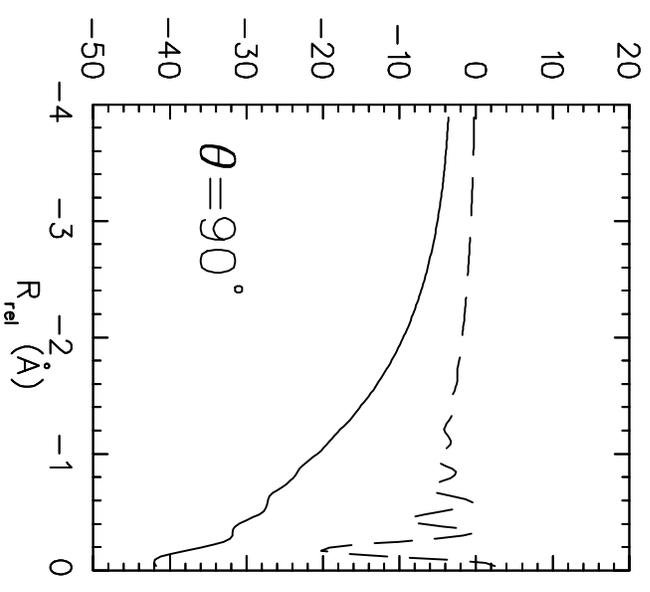
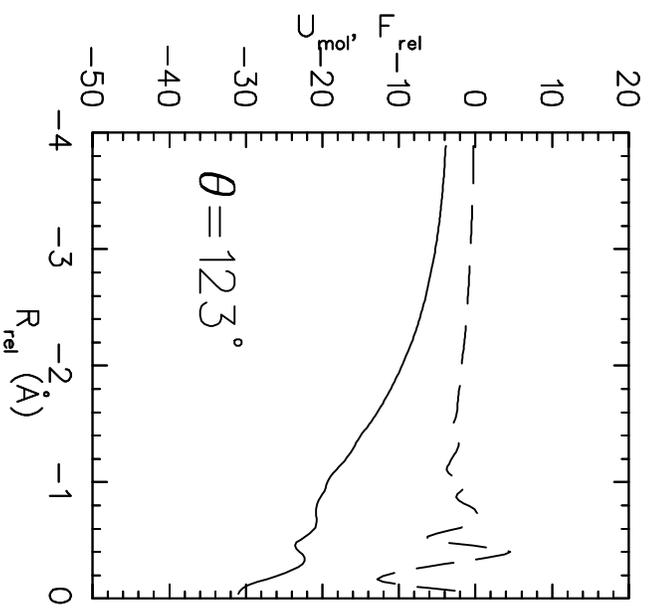
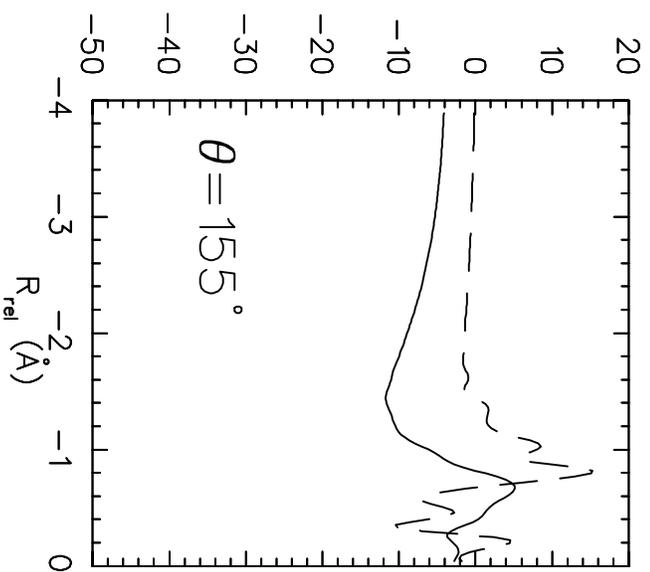
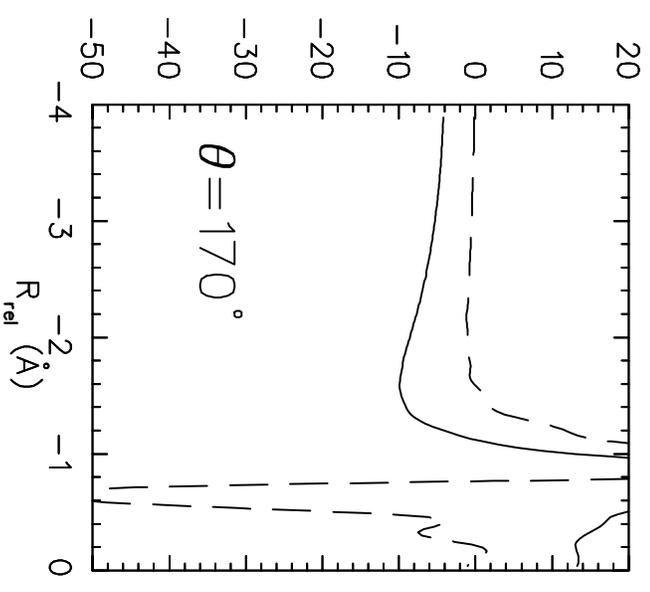
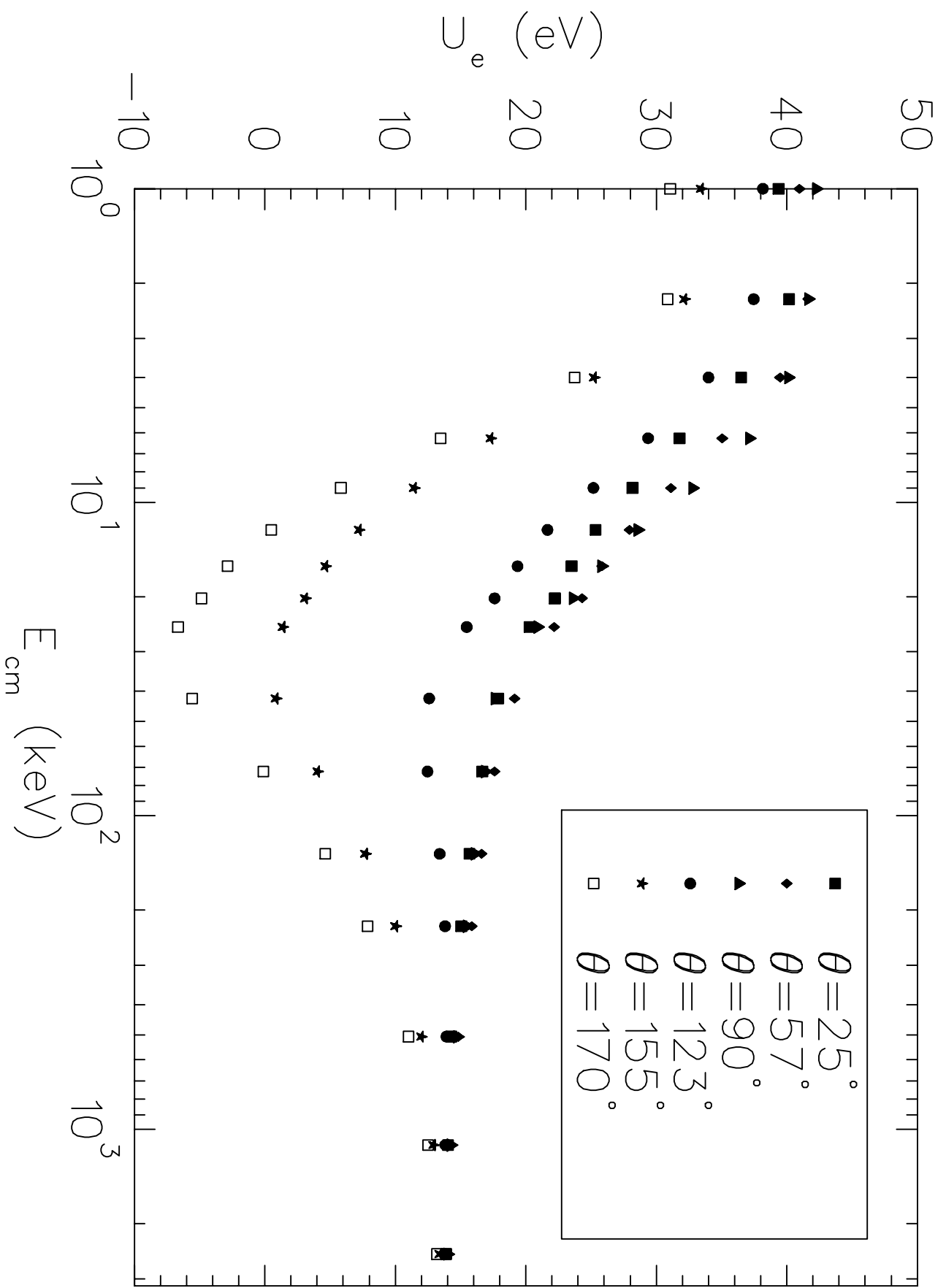
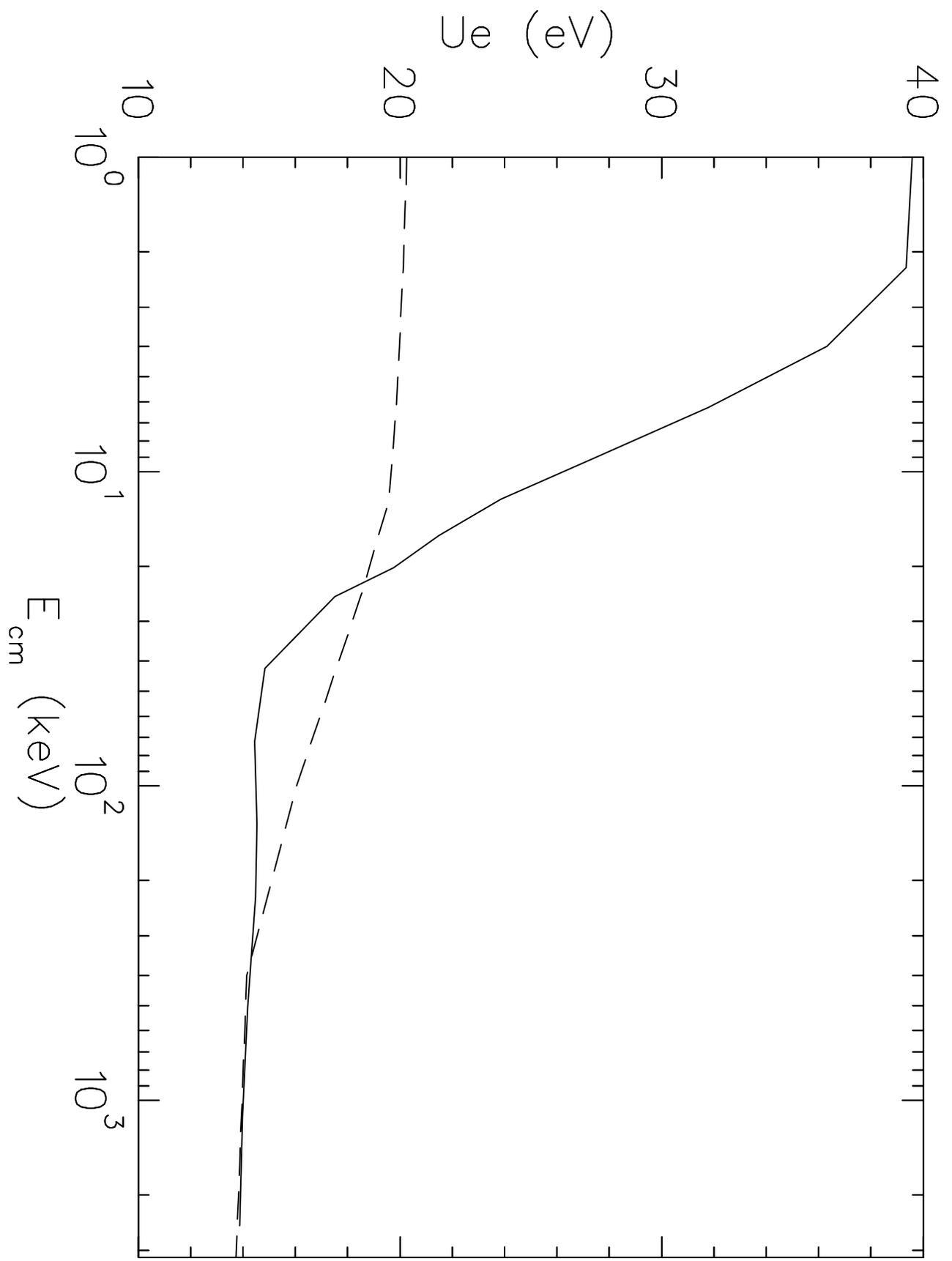
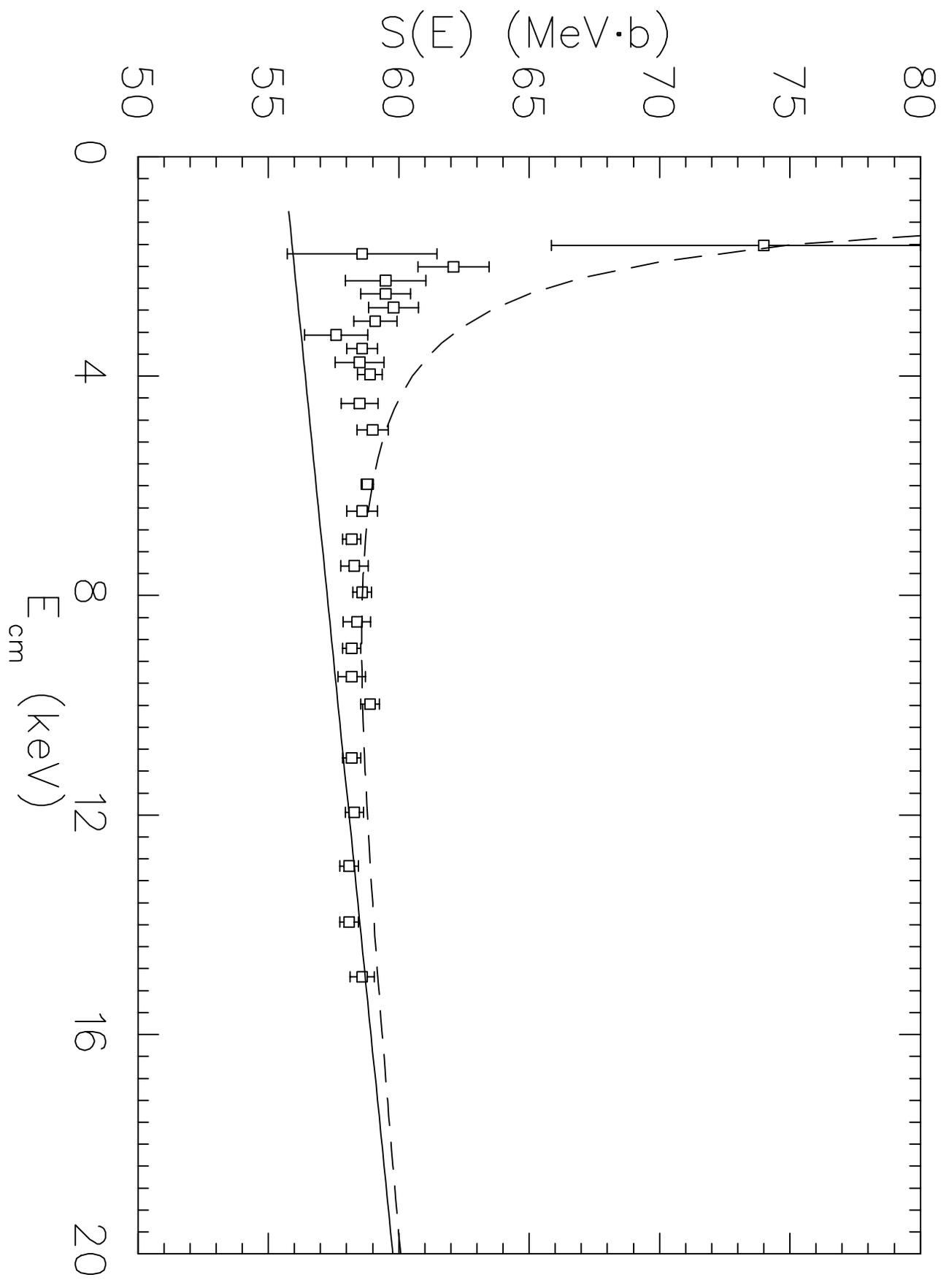
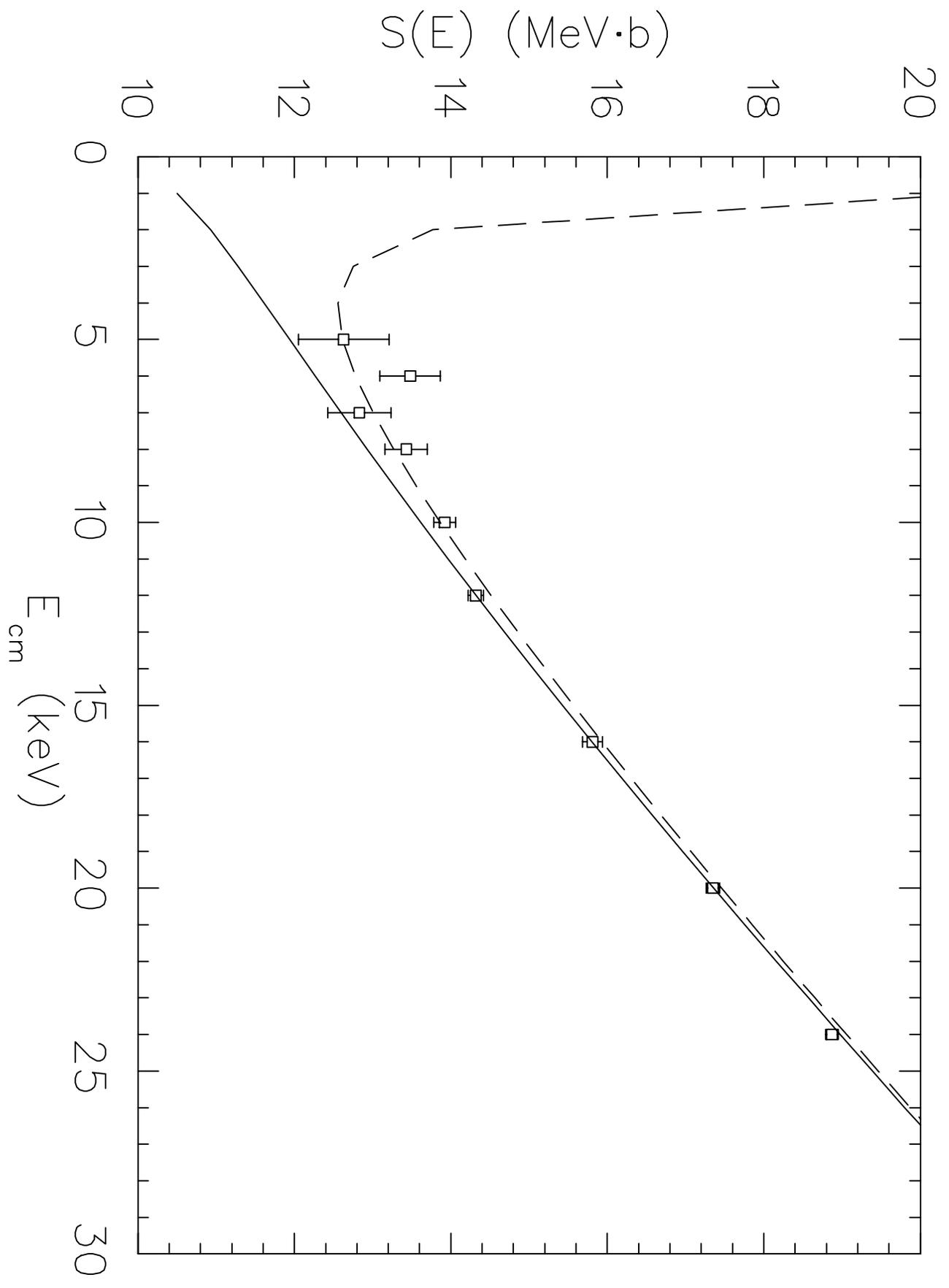
\end{document}